\documentclass[prb, preprint, twocolumn, amsmath, amssymb, reprint, superscriptaddress,]{revtex4-1}

\usepackage{mathrsfs}
\usepackage{graphicx,verbatim}
\usepackage{dcolumn}
\usepackage{bm}
\usepackage{sidecap}
\usepackage{braket}
\usepackage{color}

\graphicspath{{/user/rhaertle/epsfiles/Propplots/},{/user/rhaertle/Berlin08/},
{/user/rhaertle/epsfiles/Rate-Report/},{/user/rhaertle/epsfiles/RateCoul/},
{/user/rhaertle/epsfiles/Benesch/},{/user/rhaertle/PaperI/},{/user/rhaertle/PaperII/},
{/user/rhaertle/epsfiles/Cizek/}}

\begin{document}

\title{Extending the hierarchical quantum master equation approach to low
temperatures and realistic band structures}

\author{A. Erpenbeck}
\affiliation{
Institute for Theoretical Physics and Interdisciplinary Center 
for Molecular Materials, Friedrich-Alexander-Universit\"at Erlangen-N\"urnberg,
Staudtstr.\ 7/B2, D-91058, Germany
}
\author{C. Hertlein}
\affiliation{
Institute for Theoretical Physics and Interdisciplinary Center 
for Molecular Materials, Friedrich-Alexander-Universit\"at Erlangen-N\"urnberg,
Staudtstr.\ 7/B2, D-91058, Germany
}
\author{C. Schinabeck}
\affiliation{
Institute for Theoretical Physics and Interdisciplinary Center 
for Molecular Materials, Friedrich-Alexander-Universit\"at Erlangen-N\"urnberg,
Staudtstr.\ 7/B2, D-91058, Germany
}
\affiliation{
Institute of Physics, University of Freiburg, Hermann-Herder-Str.\ 3, D-79104 Freiburg, Germany
}
\author{M. Thoss}
\affiliation{
Institute for Theoretical Physics and Interdisciplinary Center 
for Molecular Materials, Friedrich-Alexander-Universit\"at Erlangen-N\"urnberg,
Staudtstr.\ 7/B2, D-91058, Germany
}
\affiliation{
Institute of Physics, University of Freiburg, Hermann-Herder-Str.\ 3, D-79104 Freiburg, Germany
}

\date{\today}

\begin{abstract}

The hierarchical quantum master equation (HQME) approach is an accurate method to describe quantum transport in interacting nanosystems. 
It generalizes perturbative master equation approaches by including higher-order contributions as well as non-Markovian memory and allows for the systematic convergence to the numerically exact result. As the HQME
method relies on a decomposition of the bath correlation function in terms of exponentials, however, its application to systems at low temperatures coupled to baths with complexer band structures has been a challenge.
In this publication, we outline an extension of the HQME approach, which uses a re-summation over poles and can be applied to calculate transient currents at a numerical cost that is independent of temperature and band structure of the baths. We demonstrate the performance of the extended HQME approach for noninteracting tight-binding model systems of increasing complexity as well as for the spinless Anderson-Holstein model.
\end{abstract}


\maketitle

\section{Introduction}

	Quantum transport through nanostructures, such as, e.g., molecular junctions, is an active field of research, which combines the possibility to study fundamental aspects of non-equilibrium many-body quantum physics at the nanoscale with the perspective for applications in nanoelectronic devices. \cite{Aviram1974, Nitzan2001, Joachim2205, Cuniberti, Cuevas_Scheer}
	From the theoretical side, there are several approaches capable of describing quantum transport in nanosystems.\cite{Thoss2018} Approximate methods include quantum master equations,\cite{May2002, Lehmann04, Timm08, Mitra2004, Pedersen2005, Donarini06, Zazunov06, Harbola2006, Leijnse2008, Esposito09, Kast11, Volkovich2011, Rainer2011} scattering theory,\cite{Ness01, Cizek2004, Toroker2007, Benesch2008, Jorn2009} and the non-equilibrium Green's function approach.\cite{Caroli1972, Hyldgaard1994, Galperin_Vib_Effects, Galperin2006, Rainer2008, Entin2009,Bergfield09,Novotny11, Volkovich2011, Haertle11_3, Rainer2013,Buerkle13, Haertle_13, Erpenbeck2015}
	A numerically exact treatment can, for example, be provided by means of path integrals,\cite{Muehlbacher2008,Werner09,Schiro09,Cohen13,Huetzen12,Simine13,Cohen15} multiconfigurational wave-function methods,\cite{Wang2009, Wang16} numerical renormalization-group theories,\cite{Anders06,Heidrich-Meisner2009,Eckel10,Jovchev13} a combination of reduced density matrix techniques and impurity solvers,\cite{Cohen11, Wilner2014}  and the hierarchical quantum master equations (HQME) approach (also called hierarchical equation of motion (HEOM) method).
	The latter method, which will be the focus of this paper, was originally developed by Tanimura \textit{et al.} to describe relaxation dynamics in quantum systems,\cite{Tanimura1989, Tanimura2006} and later extended by Yan \textit{et al.}\cite{Jin2008, Zheng2009, Zheng2012, Li2012, Zheng2013, Cheng2015, Ye2016} and H\"artle \textit{et al.}\cite{Haertle2013a, Haertle2013a, Haertle2014, Haertle2015, Wenderoth2016, Schinabeck2016} to study charge transport.
	
	In its original formulation, the HQME method employs a decomposition of the bath correlation function in terms of exponentials, which are also referred to as poles.\cite{Jin2008, Zheng2012, Haertle2013a} Although mathematically exact, only a finite number of exponentials can be taken into account in numerical calculations, thus effectively restricting the efficient application of the HQME method to systems at higher temperatures coupled to simple baths.
	In order to circumvent this restriction, more efficient decomposition schemes for the bath correlation function allowing for a systematic construction of a hierarchy were proposed, such as the Pade decomposition,\cite{Hu2010, Hu2011} the Chebyshev decomposition,\cite{Tian2012, Popescu2015, Popescu2016} an expansion in terms of a complete set of orthogonal functions,\cite{Tang2015} or hybrid approaches combining different decomposition schemes.\cite{Ye2017} Although these extensions improved the applicability of the HQME approach profoundly, they still rely on a decomposition that needs truncation for numerical applications, thus introducing an approximation, which limits the applicability to certain parameter regimes. In order to lift this restriction, we propose an extension of the HQME approach, which employs an analytic re-summation of the decomposition thus avoiding the truncation in numerical applications. This extends the applicability of the HQME method to quantum transport in systems at low temperatures coupled to structured baths.

	The outline of this paper is as follows: In Sec.\ \ref{sec:theory}, we establish the theory. To this end, we introduce the model system in Sec.\ \ref{sec:model} and review the HQME method in Sec.\ \ref{sec:HQME}. In Sec.\ \ref{sec:RSHQME}, we outline the extension of the HQME method, that uses a re-summation over poles. In Sec.\ \ref{sec:results} we demonstrate the performance of this extended HQME method by calculating transient currents for model systems of increasing complexity and comparing the results to traditional HQME calculations. A conclusion is given in Sec.\ \ref{sec:conclusion}.

\section{Theory}\label{sec:theory}
	\subsection{Model}\label{sec:model}
		In order to study quantum transport through nanosystems, we consider a typical model for a molecular junction, comprising the molecule (in the following referred to as `system'), which is coupled to two macroscopic leads (representing the `bath').
		The corresponding Hamiltonian is given by
		\begin{eqnarray}
			H	&=&	H_\text{S} + H_\text{L} + H_\text{R} + H_\text{SL} + H_\text{SR},
		\end{eqnarray}
		where $H_\text{S}$ describes the system, $H_\text{L/R}$ the leads and $H_\text{SL/R}$ the coupling between the system and the leads, which enables transport.
		The left and right lead is modeled as a continuum of noninteracting electronic states of energy $\epsilon_k$,
		\begin{eqnarray}
			H_\text{L/R}	&=&	\sum_{k\in\text{L/R}} \epsilon_k c_k^\dagger c_k ,
		\end{eqnarray}
		where $c_k^{\dagger} / c_k$ are the corresponding creation and annihilation operators.
		The coupling between the system and the leads is described by the Hamiltonian
		\begin{eqnarray}
			H_\text{SL/R}	&=&	\sum_{\nu\in\text{S}\atop k\in\text{L/R}} \left( V_{\nu k} c_k^\dagger d_\nu  + \text{h.c.} \right), \label{eq:Hamiltonian_coupling}
		\end{eqnarray}
		with $d_\nu^\dagger / d_\nu$ being the electronic creation and annihilation operators of the system.
		This form of the coupling between the system and the leads gives rise to the spectral density of the leads 
		\begin{eqnarray}
			\Gamma_{\text{L/R} \nu\nu'}(\epsilon) = 2\pi\sum_{k\in\text{L/R}} V_{\nu k} V_{\nu'k}^* \delta(\epsilon-\epsilon_k),
		\end{eqnarray}
		which depends on the electronic energies $\epsilon_k$ and hence incorporates the band structure of the lead.

	\subsection{HQME approach}\label{sec:HQME}
		The HQME theory provides an equation of motion for the reduced density matrix $\rho(t)$ of a system coupled to one or several baths, which are in our case the leads.
		For a system coupled to the leads via the Hamiltonian (\ref{eq:Hamiltonian_coupling}), all information about the influence of the leads on the system is encoded in the two-time bath correlation function 
		\begin{eqnarray}
			C_{l\nu\nu'}^\pm(t-t')	&=&	\frac{1}{2\pi} \int d\epsilon \ e^{\pm \frac{i}{\hbar}\epsilon (t-t')} \Gamma_{l\nu\nu'}(\epsilon) f(\pm\epsilon, \pm\mu_l) , \nonumber \\ \label{eq:correlation_func_general}
		\end{eqnarray}
		with the spectral density $\Gamma_{l\nu\nu'}(\epsilon)$ of lead $l$ and the Fermi distribution function $f(\epsilon, \mu) = \left( 1 + \exp(\beta(\epsilon-\mu))  \right)^{-1}$. Here, $\beta=\frac{1}{k_\text{B}T}$ where $k_B$ is the Boltzmann constant, $T$ the temperature and $\mu$ the chemical potential.
		In order to obtain a closed set of equations within the HQME approach, it is expedient to represent the bath correlation function as a sum over exponentials.\cite{Jin2008}
		To this end, the Fermi distribution and the  spectral density of states $\Gamma_{l\nu\nu'}(\epsilon)$ are separately represented by a sum-over-poles scheme,
		\begin{eqnarray}
			&&C_{l\nu\nu'}^\pm(t-t')	\equiv 	\sum_{q=1}^\infty     \eta_{l\nu\nu'q\pm}  e^{-\gamma_{l\nu\nu'q\pm} (t-t')}  \label{eq:decomposition_correlation_function} \\
						&&=		\sum_{p'=1}^{\infty}  \tilde\eta_{l\nu\nu'p'\pm}  e^{-\tilde\gamma_{l\nu\nu'p'\pm} (t-t')}  
							    + 	\sum_{p=1}^\infty     \check\eta_{l\nu\nu'p}  e^{-\check\gamma_{lp\pm} (t-t')}  \nonumber .
		\end{eqnarray}
		In the notation used here, the parameters $\tilde\eta_{l\nu\nu'p'\pm}$ and $\tilde\gamma_{l\nu\nu'p'\pm}$ correspond to the decomposition of $\Gamma_{l\nu\nu'}(\epsilon)$, whereas $\check\eta_{l\nu\nu'p}$ and $\check\gamma_{lp\pm}$ stem from the decomposition of the Fermi function. Common approaches for obtaining these representation are the Matsubara\cite{Mahan, Tanimura2006, Jin2008} and the Pade decomposition.\cite{Hu2010, Hu2011}
		
		For details on the derivation of the HQME for a system-bath coupling of the form Eq.\ (\ref{eq:Hamiltonian_coupling}) we refer to Refs.\ \onlinecite{Jin2008, Zheng2012, Haertle2013a}.
		The equation of motion for the $n^{\text{th}}$-tier auxiliary density operator is given by
		\begin{eqnarray}
		 	\frac{\partial}{\partial t} \rho_{j_1 \dots j_n}^{(n)}(t) 	&=& 
			\left[-\frac{i}{\hbar} \mathcal{L}_\text{S} -\left( \sum_{m=1}^n \gamma_{j_m} \right)  \right] \rho_{j_1 \dots j_n}^{(n)}(t)
			\nonumber \\&&
			-i \sum_{m=1}^n (-1)^{n-m} \mathcal{C}_{j_m} \rho_{j_1\dots j_{m-1} j_{m+1} \dots j_n}^{(n-1)}(t) \nonumber \\&&
			-\frac{i}{\hbar^2} \sum_{j} A^{\overline{\sigma_{j}}}_{\nu_j} \rho_{j_1 \dots j_n j}^{(n+1)}(t),
			\label{eq:EQM_nth_tier}
		\end{eqnarray}
		with the multi-index $j_i = (l_i, \nu_i, \nu_i', q_i, \sigma_i)$, where $l_i \in \lbrace \text{L, R} \rbrace$, $\nu_i$, $\nu_i'$ are electronic indices of the system, $\sigma_i = \pm 1$ and $q_i$ being the pole-index stemming from the decomposition of the bath correlation function in terms of exponentials. $\rho^{(0)}(t)$ is the reduced density operator of the system, the higher tier auxiliary density operators encode the influence of the leads on the system dynamics.	
		Further, $\overline{\sigma} = -\sigma$ and $\mathcal{L}_\text{S} O= [H_\text{S}, O]$. The objects $A^{{\sigma}}_{\nu}$ and $\mathcal{C}_{j}$ couple the $n^{\text{th}}$-tier to the $(n+1)^{\text{th}}$- and $(n-1)^{\text{th}}$-tier, respectively, and act upon the auxiliary density operators as
		\begin{subequations}
		\begin{eqnarray}
			A^{\sigma}_{\nu} \rho^{(n)}(t)	&=&	d^\sigma_{\nu} \rho^{(n)}(t) + (-1)^n \rho^{(n)}(t) d^\sigma_{\nu} \ , \\
			\mathcal{C}_{j} \rho^{(n)}(t)	&=&	\eta_{l\nu\nu'q\sigma}\ d^\sigma_{\nu'} \rho^{(n)}(t) - (-1)^n \eta_{l\nu\nu'q\overline\sigma}^*\ \rho^{(n)}(t) d^\sigma_{\nu'} . \nonumber \\ 
		\end{eqnarray}
		\end{subequations}
		In these equations, we have used the shorthand notation $d_{\nu}^- \equiv d_{\nu}$ and $d_{\nu}^+ \equiv d_{\nu}^\dagger$. This leads to an infinite set of coupled equations of motion. To this point, the HQME approach is exact for the Hamiltonian of the above given form and does not include any approximation. For applications, however,  the hierarchy needs to be truncated in a suitable manner.\cite{Tanimura1991, Yan2004, Xu2005, Schroeder2007} Further, only a finite number of poles $q_i$ can be used to represent the bath correlation function. The HQME approach is therefore particularly efficient for the description of systems, where a manageable number of poles represents a good approximation, which is the case for simple spectral densities at higher temperatures.
		
		Within the HQME approach, observables are represented via the reduced density matrix and the auxiliary density operators. For the study of transport through nanosystems, the electronic population of the system and the current are important observables of interest. While the population is given by the diagonal elements of the density matrix $\rho(t)$, the current for lead $l$ is given  by
		\begin{eqnarray}
			I_l(t)	&=&	\frac{ie}{\hbar^2} \sum_{\nu\nu'q} \text{Tr}_{\text{S}}\left( d_{\nu} \rho_{l\nu\nu'q+}^{(1)}(t) - d_{\nu}^\dagger \rho_{l\nu\nu'q-}^{(1)}(t) \right) \label{eq:current} ,
		\end{eqnarray}
		where Tr$_\text{S}$ denotes the trace over the system degrees of freedom.

	\subsection{Re-summation over poles}\label{sec:RSHQME}
		In order to extend the HQME formalism, we introduce the following weighted sum over poles
		\begin{eqnarray}
		      \mathcal{R}_{a_1 \dots a_n}^{(n)} (t, t_1, \dots, t_n)	&=&	\sum_{q_1 \dots q_n}	\prod_{m=1}^n  e^{\gamma_{j_m}(t-t_m)} \rho_{j_1 \dots j_n}^{(n)}(t), \nonumber \\ \label{eq:def_R}
		\end{eqnarray}
		\\
		\newline
		such that an infinite number of poles is treated within one object. Here, we use the multi-index $a_i = (l_i, \nu_i, \nu_i',\sigma_i)$, with $l_i \in \lbrace \text{L, R} \rbrace$, $\nu_i$, $\nu_i'$ electronic indices of the system and $\sigma_i = \pm 1$, which does not have a pole-index. As this extension of the HQME approach relies on a re-summation over poles, we will abbreviate it as RSHQME in the following. Taking the derivative of Eq.\ (\ref{eq:def_R}) with respect to time $t$ and using Eq.\ (\ref{eq:EQM_nth_tier}), we arrive at the equation of motion
		\begin{widetext}
		\begin{eqnarray}
				 	\frac{\partial}{\partial t} \mathcal{R}_{a_1 \dots a_n}^{(n)} (t, t_1, \dots, t_n) &=&
			\label{eq:RS_EQM_nth_tier} 
			-\frac{i}{\hbar} \mathcal{L}_\text{S} \mathcal{R}_{a_1 \dots a_n}^{(n)} (t, t_1, \dots, t_n) 
			-\frac{i}{\hbar^2} \sum_{a} A^{\overline{\sigma_{a}}}_{\nu_a} \mathcal{R}_{a_1 \dots a_n a}^{(n+1)}(t, t_1 \dots, t_n, t)
			\nonumber \\&&
			-i \sum_{m=1}^n \mathscr{C}_{a_m}(t, t_m) \mathcal{R}_{a_1\dots a_{m-1} a_{m+1} \dots a_n}^{(n-1)}(t, t_1, \dots, t_{m-1}, t_{m+1}, \dots, t_{n}), 
		\end{eqnarray}
		where we have defined
		\begin{eqnarray}
			\mathscr{C}_{a_m}(t, t_m) \mathcal{R}^{(n)}	&=&	(-1)^{n+1-m} \left( \xi_{a_m}(t, t_m)  d_{\nu_m'}^\sigma \mathcal{R}^{(n)}  - (-1)^n \tilde\xi_{a_m}(t, t_m)  \mathcal{R}^{(n)} d_{\nu_m'}^\sigma \right)    , \label{eq:mathscr_C}
		\end{eqnarray}
		\end{widetext}
		with
		\begin{subequations}
		\begin{eqnarray}
			\xi_{a}(t, t_m)	&=&	\sum_{q=1}^\infty     \eta_{l\nu\nu'q\pm}\  e^{-\gamma_{l\nu\nu'q\pm} (t-t_m)}
			\label{eq:def_xi} ,\\
			\tilde\xi_{a}(t, t_m)	&=&	\sum_{q=1}^\infty     \eta_{l\nu\nu'q\mp}^*\  e^{-\gamma_{l\nu\nu'q\pm} (t-t_m)} .
		\end{eqnarray}
		\end{subequations}
		The only remainder of the decomposition of the bath correlation function in terms of exponentials are the sums $\xi_{a}$ and $\tilde\xi_{a}$. Notice that $\xi_{a}$ is indeed the bath correlation function as becomes apparent upon comparing Eqs.\ (\ref{eq:decomposition_correlation_function}) and (\ref{eq:def_xi}). For exact results, these sums need to be evaluated analytically. Approximate results, can be obtained by numerically evaluating the sums, where several thousand poles can easily be included.
		
		Considering the $\mathcal{R}^{(n)}$ where all time arguments are equal, the previous auxiliary density operators, summed over all poles $\sum_{q_1 \dots q_n}$, are recovered, which are important for the representation of observables. Of particular interest for this work is the electronic current, which is given by
		\begin{eqnarray}
			I_l(t)	&=&	\frac{ie}{\hbar^2} \sum_{\nu\nu'} \text{Tr}_{S}\left( d_{\nu} \mathcal{R}_{l\nu\nu'+}^{(1)}(t,t) - d^\dagger_{\nu} \mathcal{R}_{l\nu\nu'-}^{(1)}(t,t) \right)  .
			\nonumber \\ \label{eq:current_RS_HQME}
		\end{eqnarray}
		
		Although the RSHQME method is an extension of the HQME approach that can deal with the decomposition of the bath correlation function in an exact way, it comes with a numerical drawback. As the objects $\mathcal{R}^{(n)}$ depend on $(n+1)$ time-arguments, the approach scales as $\mathcal{O}(t^{n+1})$ given that $n$ is the maximal tier considered. Thus, the RSHQME method is best suited to simulate the dynamics of a system for short to intermediate times. The HQME method, on the other hand, scales as $\mathcal{O}(t \cdot q^{n})$, where $q$ is the total number of poles taken into account. This restricts the HQME approach to systems that are well approximated by a small number of poles. Some numerical details of the RSHQME method are given in Appendix \ref{sec:appendix}.

	\section{Illustrative applications}\label{sec:results}
		In this section, we apply the RSHQME method to representative model systems of increasing complexity.
		To this end, we explicitly consider the simple but generic model system consisting of a single electronic state of energy $\epsilon_0$ coupled to one vibrational mode of frequency $\Omega$ as described by the system Hamiltonian
		\begin{eqnarray}
				H_\text{S}	&=&	  \epsilon_0 d^\dagger d
							+ \hbar\Omega a^\dagger a
							+ \lambda (a^\dagger +  a) d^\dagger d . \label{eq:Hamiltionian_System}
		\end{eqnarray}
		Here, $d/d^{\dagger}$ denote the electronic and $a/a^{\dagger}$ the vibrational creation and annihilation operators, respectively. $\lambda$ is the electronic-vibrational coupling strength. For $\lambda=0$, we recover a purely electronic transport problem, to which we refer as "noninteracting".
		In the following, we consider the model system with the parameters $\epsilon_0 = 0.2$eV, $\Gamma_\text{L} = \Gamma_\text{R} = \Gamma = 0.06$eV, which are representative for molecular junctions.\cite{Elbing05, Benesch2008, Benesch2009, Ballmann2010, Arroyo2010, Ballmann2012, Erpenbeck2015, Erpenbeck2016, Schinabeck2016} 
		The bias, defined as the difference between the chemical potentials of the left and the right lead, is assumed to drop symmetrically, $\mu_\text{L} = -\mu_\text{R}$.  The temperature of the system is $T=0.1$K.
		For the numerical simulations, we assumed that the total density matrix factorizes at time $t = 0$. Further, at $t=0$ the molecular electronic state is unpopulated, the vibrational mode is in its ground state, and the leads are in their thermal state.
		
		We start with a noninteracting system ($\lambda=0$) within the wide-band approximation. In order to establish the applicability of the RSHQME method, we compare its results to the outcome of the traditional HQME approach. These results provide the possibility to benchmark HQME calculations including different numbers of poles. Subsequently, we consider a model for the leads beyond the wide-band limit, demonstrating that the RSHQME method can describe the influence of band structure effects on electronic transport. Finally, we show that the RSHQME is applicable to interacting systems, thus being a numerically affordable extension of the exact HQME method to low temperatures and complex band structures. 
		To this end, we consider the spinless Anderson-Holstein model ($\lambda\neq 0$), which serves as an example for an interacting model.

	\subsection{Noninteracting system attached to leads in the wide-band limit}\label{sec:wbl}
		We first consider a noninteracting model described by the system Hamiltonian Eq.\ (\ref{eq:Hamiltionian_System}) with $\lambda=0$, coupled to leads modeled in the  wide-band limit, where the energy-dependence of the spectral density is neglected. 
		
		In the wide-band limit, it is not feasible to perform a decomposition of the constant $\Gamma_\text{L/R}(\epsilon)$ in terms of exponentials. It is more efficient to drop the first sum in the second line of Eq.\ (\ref{eq:decomposition_correlation_function}) and instead include one additional auxiliary object in every tier of the hierarchy, which is calculated as 
		\begin{subequations}
		\begin{eqnarray}
		 \tilde\rho_{(l, \sigma) j_1 \dots j_n}^{(n+1)} 	&=&	- \frac{i\hbar \Gamma}{4} 
									 \cdot \left\lbrace d^{\sigma}, \rho_{j_1 \dots j_n}^{(n)} \right\rbrace_{(-1)^{n+1}} , \\
		 \mathcal{\tilde R}_{(l, \sigma)  a_1 \dots a_n}^{(n+1)} 	&=&	- \frac{i\hbar \Gamma}{4} 
									 \cdot \left\lbrace d^{\sigma}, \mathcal{R}_{a_1 \dots a_n}^{(n)} \right\rbrace_{(-1)^{n+1}} ,
		\end{eqnarray}
		\end{subequations}
		and not by forward propagation. As this is a technical aspect which is identical for the HQME and the RSHQME method, we refrain from giving details here and refer the reader to Refs.\ \onlinecite{Croy2009, Zheng2010, Zhang2013,Kwok2014, Erpenbeck2018} for more information.
		
		For the RSHQME method, we represent the Fermi-distribution employing the Matsubara decomposition, yielding
		\begin{subequations}
		\begin{eqnarray}
			\gamma_{lp\pm}	&=&	\frac{\pi}{\hbar\beta}(2p-1) \mp \frac{i}{\hbar}\mu_l, \\
			\eta_{lp}	&=&	-\frac{i\Gamma}{\beta} .
		\end{eqnarray}
		\end{subequations}
		With these expressions for $\gamma_{lp\pm}$ and $\eta_{lp}$, we obtain
		\begin{eqnarray}
			\xi_{a_m}(t, t_m)	&=&	-\frac{i \Gamma}{\beta} \cdot  \frac{e^{\left(\frac{3\pi}{\hbar\beta} - \sigma_m \frac{i}{\hbar}\mu_l\right)(t-t_m)}}{1-e^{\frac{2\pi}{\hbar\beta}(t-t_m)}} = -\tilde\xi_{a_m}(t, t_m). \nonumber \\ \label{eq:WBL_xi}
		\end{eqnarray}
		Using Eq.\ (\ref{eq:WBL_xi}) in Eq.\ (\ref{eq:RS_EQM_nth_tier}), we arrive at a closed set of equations for the re-summed auxiliary density operators $\mathcal{R}^{(n)}$. 
		
		\begin{figure}[b!]
			\includegraphics[width=\linewidth]{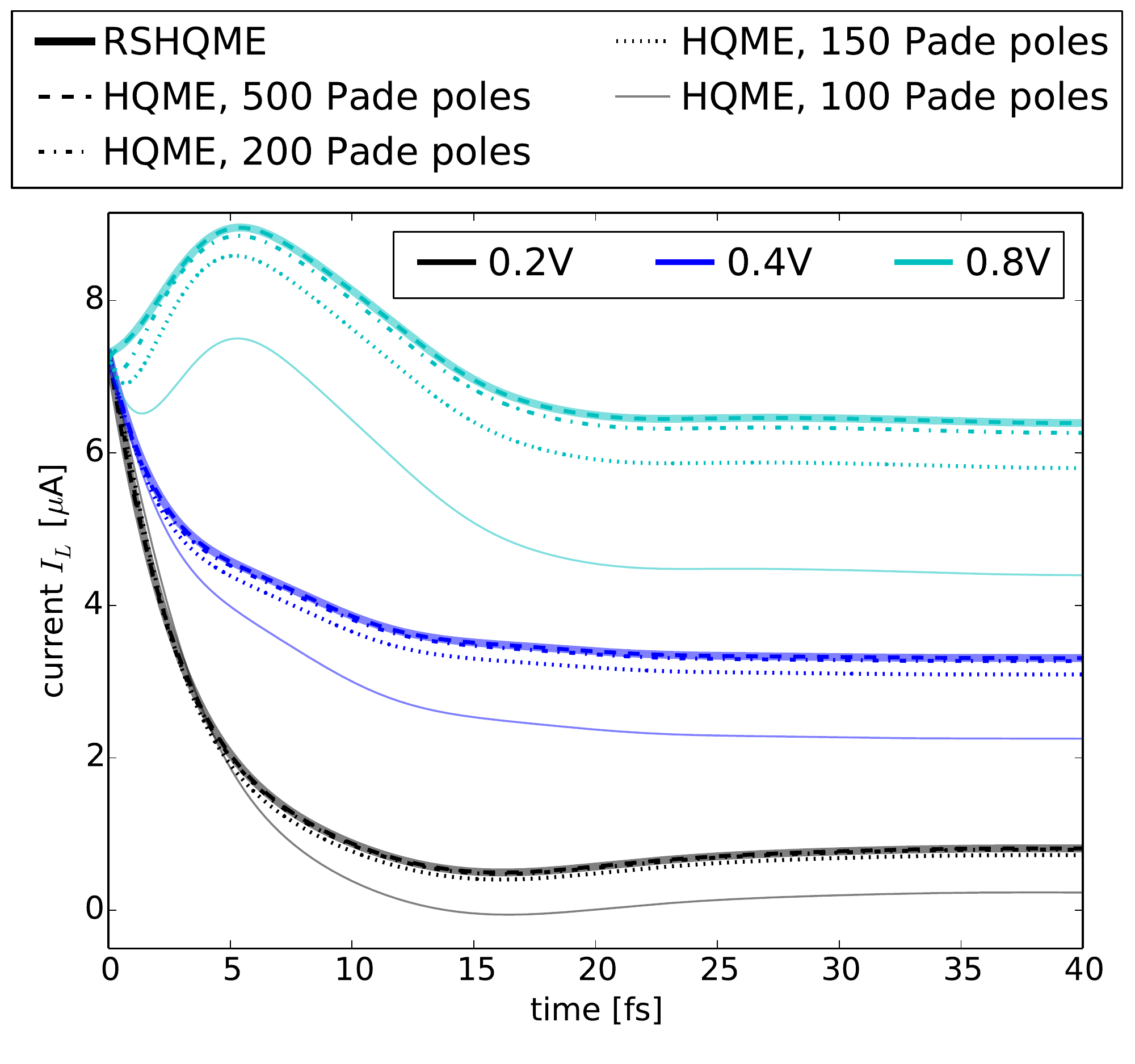}
			\caption{Current as a function of time for an applied bias voltage of $0.2$V, $0.4$V and $0.8$V. The different colors corresponds to the different bias voltages, the line style represents the numerical method used to calculate the current.}
			\label{fig:current_WBL}
		\end{figure}
		For the HQME method, we apply the Pade spectral decomposition considering $100$, $150$, $200$ and $500$ poles in the numerical calculation. Notice that the number of poles considered here is at the limit of numerical accessibility. Usually, HQME calculations take tenths of poles into account.\cite{Hu2011, Xie2012, Haertle2013a, Haertle2015, Schinabeck2016, Ye2017, Schinabeck2018} We refrain from discussing results obtained using the Matsubara decomposition for a finite number of poles, as even taking into account $500$ Matsubara poles fails to provide a reasonable result for any bias voltage. 
		
		It is noted that for a noninteracting system, the hierarchy terminates at the $2$nd-tier.\cite{Jin2008, Karlstrom2013} In the wide-band limit, it suffices to only include the $1$st-tier auxiliary density matrices for exact results for single-particle observables such as the current.\cite{Croy2009, Zheng2010, Kwok2014}

		Fig.\ \ref{fig:current_WBL} depicts the current as a function of time for different bias voltages and different numerical routines. 
		For $0.2$V, the system is in the non-resonant transport regime, whereas for $0.8$V it is in the resonant regime. For $0.4$V, the chemical potential of the left lead $\mu_{\text{L}}$ is in resonance with the molecular electronic state $\epsilon_0$. All currents show a transient behavior before approaching the steady-state value. Both the transient behavior and the steady state value of the current depend on the numerical details of the different methods. The transient behavior has already been studied in detail elsewhere.\cite{Schmidt2008, Wang2009, Wang2013, Wilner2014, Wang2016_2}
		
		We compare the current obtained by the different numerical methods. Generally, the quality of the spectral decomposition increases with an increasing number of poles,\cite{Hu2010, Hu2011}
		thus the results calculated by the HQME method converge towards the RSHQME results with increasing number of poles. For $0.2$V, which is representative for the non-resonant transport regime, the HQME calculation accounting for $200$ Pade poles performs reasonably well, whereas the results for the resonant transport regime at $0.8$V still exhibit deviations from the converged current. This is due to the fact that for the resonant transport regime, a larger part of the leads' structure in energy space has to be represented adequately, leading to an increased requirement in the number of poles taken into account with increased bias.\cite{Hu2010}
		The HQME approach taking into account $500$ Pade poles gives the same results as the RSHQME method. Only for very short times below $\sim0.3$fs, we find a small deviation between the HQME and the RSHQME result in the resonant transport regime. This can be explained by the fact that the short-time dynamics depends on a large energy span of the leads' structure.\cite{Croy2009}

	\subsection{Noninteracting system attached to leads with a finite bandwidth}\label{sec:box}
		Next, we go beyond the wide-band description of the leads, demonstrating the capability to include band-effects in the RSHQME approach. 
		Thereby, the advantage of using the RSHQME method is that the numerical cost does not depend on the band structure. Using the HQME method, on the other hand, the numerical cost depends on the model for the leads. Representing leads with a band structure increases the number of poles that need to be taken into account for the calculation, thus increasing the numerical effort.
		
\begin{figure}[tb!]
			\includegraphics[width=\linewidth]{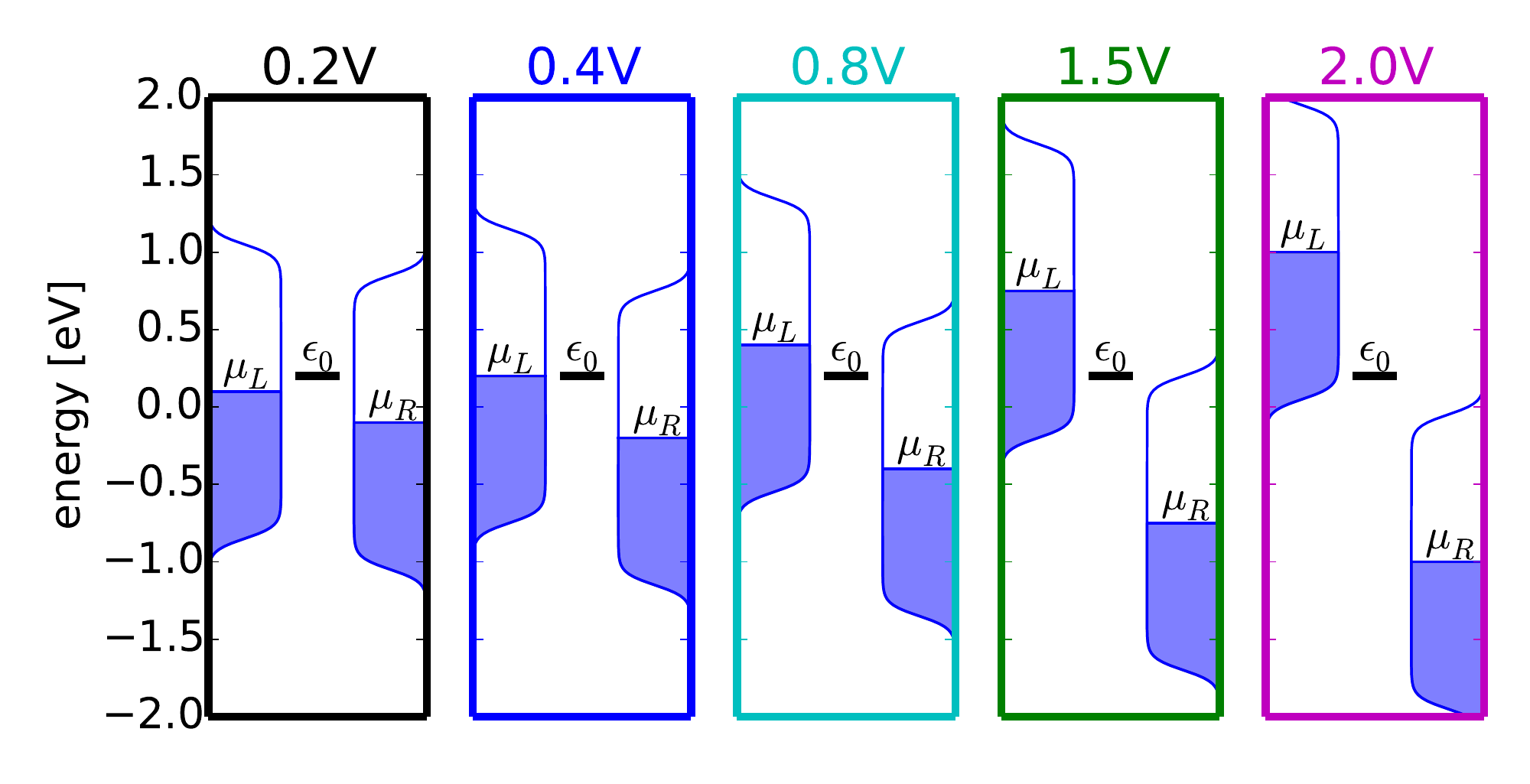}
			\includegraphics[width=\linewidth]{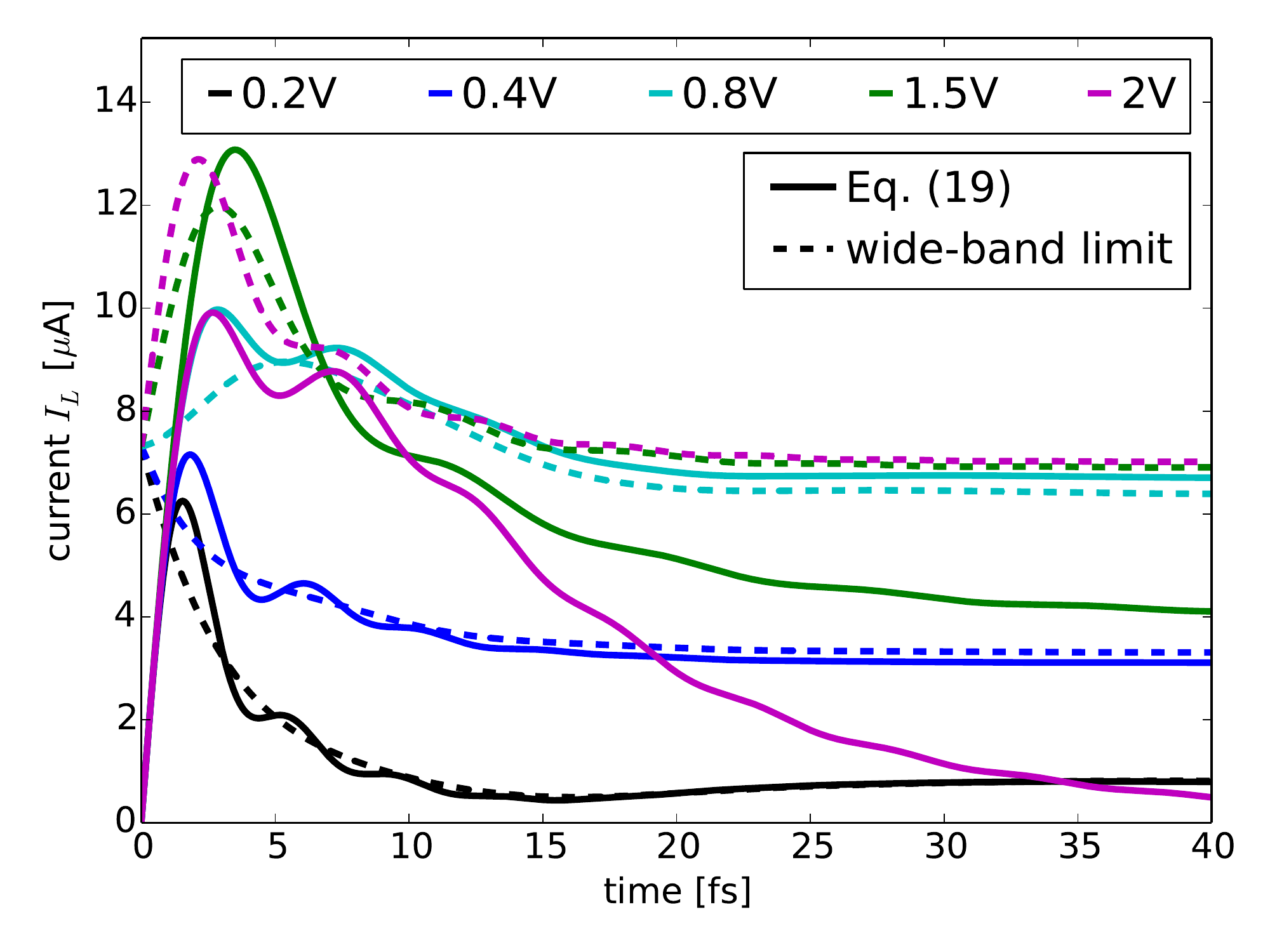}
			\caption{
			Top: Energy level scheme of the molecule and the leads as a function of applied bias voltage.
			Bottom: Current as a function of time for an applied bias voltage of $0.2$V---$2.0$V. The dashed lines depict the current in the wide-band limit, the solid lines are obtained for a system coupled to leads with a finite bandwidth. The colors represent different applied bias voltages.}
			\label{fig:bandstructure}
		\end{figure}

		We consider the noninteracting model described by the Hamiltonian Eq.\ (\ref{eq:Hamiltionian_System}) with $\lambda=0$, the spectral density of the leads is modeled by a box function with smooth edges as described by the function
		\begin{eqnarray}
			\Gamma_{l, \mu_l}(\epsilon)	&=&	\frac{\Gamma}{
									\left( 1 + e^{\alpha  (\epsilon - \epsilon_C - \mu_l} \right)
								\cdot   \left( 1 + e^{-\alpha (\epsilon + \epsilon_C - \mu_l} \right)
								} \label{eq:bandstructure}
		\end{eqnarray}
		with $\alpha = 25$eV$^{-1}$ and $\epsilon_C = 0.95$eV. This form of $\Gamma_{l, \mu_l}(\epsilon)$ is also referred to as wide-band limit with soft cut-off and was already used by \citet{Schmidt2008} who investigated the influence of the band structure on the transient current or, in a different context, by \citet{Muehlbacher2008}. The parameters $\alpha$ and $\epsilon_C$ where chosen such that for low bias voltages, the system behaves like in the wide-band case, however for higher bias voltages, band edge effects become important. The bias voltage is assumed to shift the energy levels of the leads. $\Gamma_{l, \mu_l}(\epsilon)$ as a function of applied bias voltage is depicted in Fig.\ \ref{fig:bandstructure} (top).

		In order to describe this non-trivial spectral density within the RSHQME methodology, we represent Eq.\ (\ref{eq:bandstructure}) by a modified Pade decomposition
		\begin{eqnarray}
			\Gamma_{l, \mu_l}(\epsilon) 	&=&	\frac{Q_{l,\mu_l}(\epsilon)}{P_{l,\mu_l}(\epsilon)} 	\approx 	\frac{\sum_{n=0}^{N-1} q_{n} \epsilon^n}{1 + \sum_{n=1}^{N} p_{n} \epsilon^n}. \label{eq:representation}
		\end{eqnarray}
		Sampling $\Gamma_{l, \mu_l}(\epsilon)$ by $2N$ $\epsilon$-values leads to a set of coupled linear equations, which can be solved for $q_{n}$ and $p_{n}$. 
		Applying the Bairstow algorithm,\cite{McNamee} we determine the roots of $P_{l,\mu_l}(\epsilon)$ which are used to perform the energy integral in Eq.\ (\ref{eq:correlation_func_general}) using Cauchy's residual theorem. In this way, we obtain the values for $\tilde\eta_{lp'\pm}$ and $\tilde\gamma_{lp'\pm}$. However, the details of the decomposition of $\Gamma_{l, \mu_l}(\epsilon)$ are not of importance, because the RSHQME approach works with any decomposition scheme. As only the sum over poles enters the equation of motion, any number of poles $N$ can be included in the calculation. Thus the RSHQME method opens up the possibility to accurately describe leads of arbitrary complexity. The approach of fitting complex band structures in order to represent band effects was already discussed in Refs.\ \onlinecite{Meier1999, Welack2006} and was for example used in explicit calculations by \citet{Xie2012}, who expanded the density of states of 1D tight-binding chains in terms of Lorentzians and therefrom performed HQME calculations.
		
		Fig.\ \ref{fig:bandstructure} (bottom) compares the time-dependent current for a system attached to leads with different band structures. The solid lines represent the current for different bias voltages, where the leads are modeled by Eq.\ (\ref{eq:bandstructure}). The associated band structure as a function of bias voltage is given in Fig.\ \ref{fig:bandstructure} (top). The dashed lines in Fig.\ \ref{fig:bandstructure} (bottom) correspond to the results obtained in the wide-band limit already discussed in Sec.\ \ref{sec:wbl}.
		
		The most striking difference between the results for the two lead models is observed for the current at short times. The fact that the wide-band description of the leads results in an instantaneous finite current, which is unphysical, was already studied by \citet{Schmidt2008}. Furthermore, the precise form of the transient current also depends on the details of the lead band structure. For bias voltages $0.2$V, $0.4$V and $0.8$V, the current for the system attached to leads with a finite band approaches after a certain time a value similar to that obtained in the wide-band limit. 
		For the larger bias voltages $1.5$V and $2$V, band edge effects become important also in the steady state. The long-time current is smaller than in the wide-band limit because the density of states in the leads in resonance with the molecular electronic states is decreased with bias.

	\subsection{Interacting model system}\label{sec:el-vib}
		To demonstrate the applicability of the RSHQME approach to interacting systems, we consider the spinless Anderson-Holstein model given by the Hamiltonian Eq.\ (\ref{eq:Hamiltionian_System}), where we set the energy of the vibrational mode to $\hbar\Omega=0.2$eV. We consider two different coupling strengths, $\lambda=0.05$eV and $\lambda=0.1$eV, and compare the results to the noninteracting system $\lambda=0$.
		These model parameters are in the typical range for molecular junctions, similar parameters have been used in our recent work on the spinless Anderson-Holstein model.\cite{Schinabeck2016}
		Notice that we consider a system in the anti-adiabatic regime, $\hbar\Omega > \Gamma$, as the influence of the electronic-vibrational interaction on the short-time dynamics is most pronounced under these conditions.
		The leads are modeled in the wide-band limit.
		\begin{figure}[tb!]
			\includegraphics[width=\linewidth]{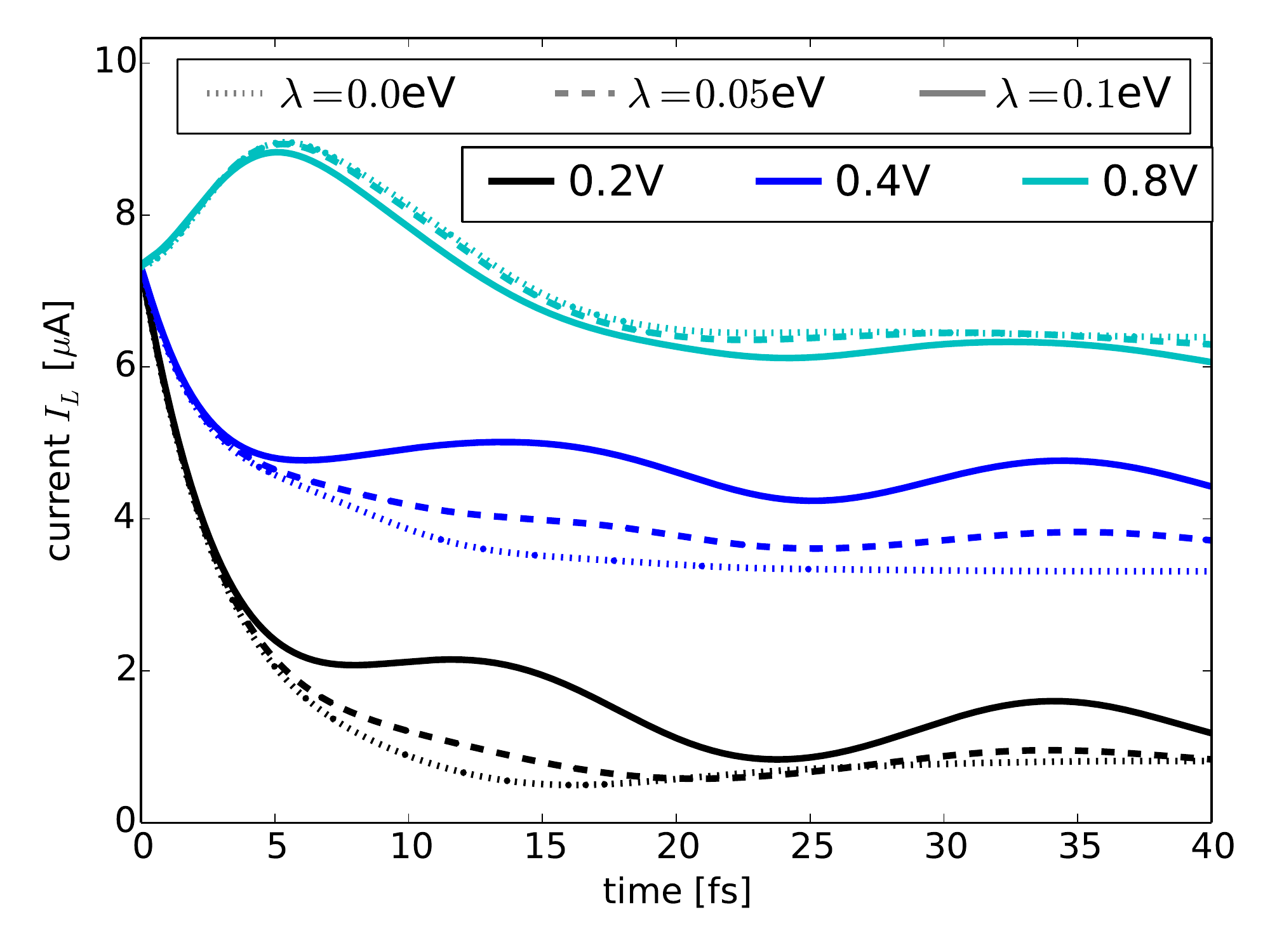}
			\caption{
			Current as a function of time for an applied bias voltage of $0.2$V, $0.4$V and $0.8$V. The different colors correspond to the different bias voltages, the line style represents the electronic-vibrational interaction strength $\lambda=0$eV, $0.05$eV, $0.1$eV.
			}
			\label{fig:AH}
		\end{figure}
		Fig.\ \ref{fig:AH} compares the current for different coupling strengths, $\lambda=0$eV, $0.05$eV, $0.1$eV, obtained by the RSHQME method. In the numerical calculations, we used three vibrational basis states for $\lambda=0.05$eV and six vibrational basis states for $\lambda=0.1$eV. The finding that a relatively small vibrational basis set is sufficient to obtain converged results for the current is due to the comparably weak coupling strength $\lambda$, leading to a small non-equilibrium vibrational excitation, and the fact that the current is not very sensitive with respect to the vibrational basis for the parameters studied here. 
		Note also that typically a smaller number of vibrational basis states is necessary to obtain converged results for short to intermediate times. For the chosen parameters, the current was converged employing the hierarchy up to the $2$nd-tier. Converged HQME calculations would require about $500$ Pade poles, which is numerically very demanding and demonstrates the improvement obtained with the RSHQME method.

		The results in Fig.\ \ref{fig:AH} have been obtained for a range of bias voltages, $0.2$V, $0.4$V, and $0.8$V, which span a variety of different transport processes. The current at $0.2$V is predominantly composed of non-resonant elastic and inelastic co-tunneling processes. For the bias voltages $0.4$V and $0.8$V, elastic and inelastic resonant transport processes are of importance. Thereby, at $0.8$V, more resonant inelastic transport processes are energetically possible than for $0.4$V. A detailed analysis of transient currents through a system with electronic-vibrational interaction has been given in Ref.\ \onlinecite{Riwar2009}.

\section{Conclusion}\label{sec:conclusion}
	In this paper, we have proposed an extension of the HQME method, which avoids the limitations imposed by the decomposition of the bath correlation function inherent to the construction of the hierarchy. In contrast to other extensions, the methodology proposed here relies on an analytic re-summation over poles rather than a more efficient parametrization of the bath correlation function, thus circumventing the shortcomings of the traditional HQME approach with respect to bath parametrization.
	
	In order to demonstrate the performance of the novel RSHQME method, we have applied the approach to model systems of increasing complexity including both interacting and noninteracting systems. The results show that the RSHQME method is able to reproduce the outcome of traditional HQME calculations for parameters where the latter can be converged with respect to the number of poles. Furthermore, we applied the RSHQME method to systems, where traditional HQME calculations are numerically prohibitively expensive. 
	
	Despite of the appeal of the newly introduced RSHQME method, numerical calculations become increasingly expensive with simulation time, which confines the applications of the RSHQME method in its current formulation to simulations for short or intermediate times. Similar as in other time-dependent density-matrix schemes, this limitation may be overcome by exploiting the fact that for realistic systems the bath correlation function decays in time.

\section*{ACKNOWLEDGEMENT}
	We thank R. H\"artle for helpful discussions.
	This work was supported by the German Research Foundation (DFG) through SFB 953 and a research grant.

\appendix
\section{Propagation scheme of the RSHQME method}\label{sec:appendix}
	In this appendix, we give some details on the propagation of the coupled differential equations (\ref{eq:RS_EQM_nth_tier}), which are the basis of the RSHQME method. 
	To this end, we exemplify how the lowest order auxiliary density matrices are calculated. A generalization to the calculation of higher tier auxiliary density operators is straightforward.\\
	
	In this work, the density matrix $\rho(t)=\mathcal{R}^{(0)}(t)$ for a time $t$ is calculated via propagation, which corresponds to integrating its differential equation
	\begin{eqnarray}
		\rho(t)		&=&	\int_0^t \ \partial_\tau \rho(\tau) \ d\tau \\
				&=&	\int_0^t \ F_0\Big( \ \rho(\tau), \ \mathcal{R}_{a_1}^{(1)}(\tau, \tau) \ \Big) \ d\tau . \nonumber
	\end{eqnarray}
	Here, $F_0$ indicates that according to Eq.\ (\ref{eq:RS_EQM_nth_tier}), the time-derivative $\partial_\tau \rho(\tau)$ is a function of $\rho(\tau)$ and $\mathcal{R}_{a_1}^{(1)}(\tau, \tau)$. Consequently, in order to calculate $\rho(t)$, the density matrix $\rho(\tau)$ needs to be known at all previous times $\tau$, which is trivial when using propagation, and also $\mathcal{R}_{a_1}^{(1)}(\tau, \tau)$ need to be calculated for every $\tau<t$.
	The latter is done by integrating the corresponding differential equation (\ref{eq:RS_EQM_nth_tier}) as
	\begin{eqnarray}
		\mathcal{R}_{a_1}^{(1)}(\tau, \tau)		&=&	\int_0^\tau \partial_{\tau_1} \mathcal{R}_{a_1}^{(1)}(\tau_1, \tau) \ d\tau_1 \\
								&=&	\int_0^\tau F_1\Big( \rho(\tau_1), \mathcal{R}_{a_1}^{(1)}(\tau_1, \tau), \mathcal{R}_{a_1 a_2}^{(2)}(\tau_1, \tau, \tau_1) \Big) d\tau_1 .\nonumber 
	\end{eqnarray}
	$F_1$ indicates that $\partial_{\tau_1} \mathcal{R}_{a_1}^{(1)}(\tau_1, \tau)$ is a function of $\rho(\tau_1)$, $\mathcal{R}_{a_1}^{(1)}(\tau_1, \tau)$ and $\mathcal{R}_{a_1 a_2}^{(2)}(\tau_1, \tau, \tau_1)$ for $\tau_1<\tau<t$.
	Notice, that only the first time argument, which corresponds to the actual physical time, is propagated. The later time argument is a parameter that is kept constant during this propagation.
	The necessary $2$nd-tier auxiliary density operators are then calculated as 
	\begin{eqnarray}
		\mathcal{R}_{a_1 a_2}^{(2)}(\tau_1, \tau, \tau_1)		&=&	\int_0^{\tau_1} \partial_{\tau_2} \mathcal{R}_{a_1 a_2}^{(2)}(\tau_2, \tau, \tau_1) \ d\tau_2 \\
										&=&	\int_0^{\tau_1} F_2\Big( \mathcal{R}_{a_1}^{(1)}(\tau_2, \tau_1), \mathcal{R}_{a_1}^{(1)}(\tau_2, \tau), \nonumber\\
										&&	\mathcal{R}_{a_1 a_2}^{(2)}(\tau_2, \tau, \tau_1), \mathcal{R}_{a_1 a_2 a_3}^{(3)}(\tau_2, \tau, \tau_1, \tau_2) \Big) d\tau_2 \nonumber
	\end{eqnarray}
	with $\tau_2<\tau_1<\tau<t$. Again, only the first time argument is propagated, whereas the others are kept constant throughout the propagation. This scheme straightforwardly generalizes to higher tiers.

	\begin{figure}[htb!]
		\centering
		\includegraphics[width=0.5\textwidth]{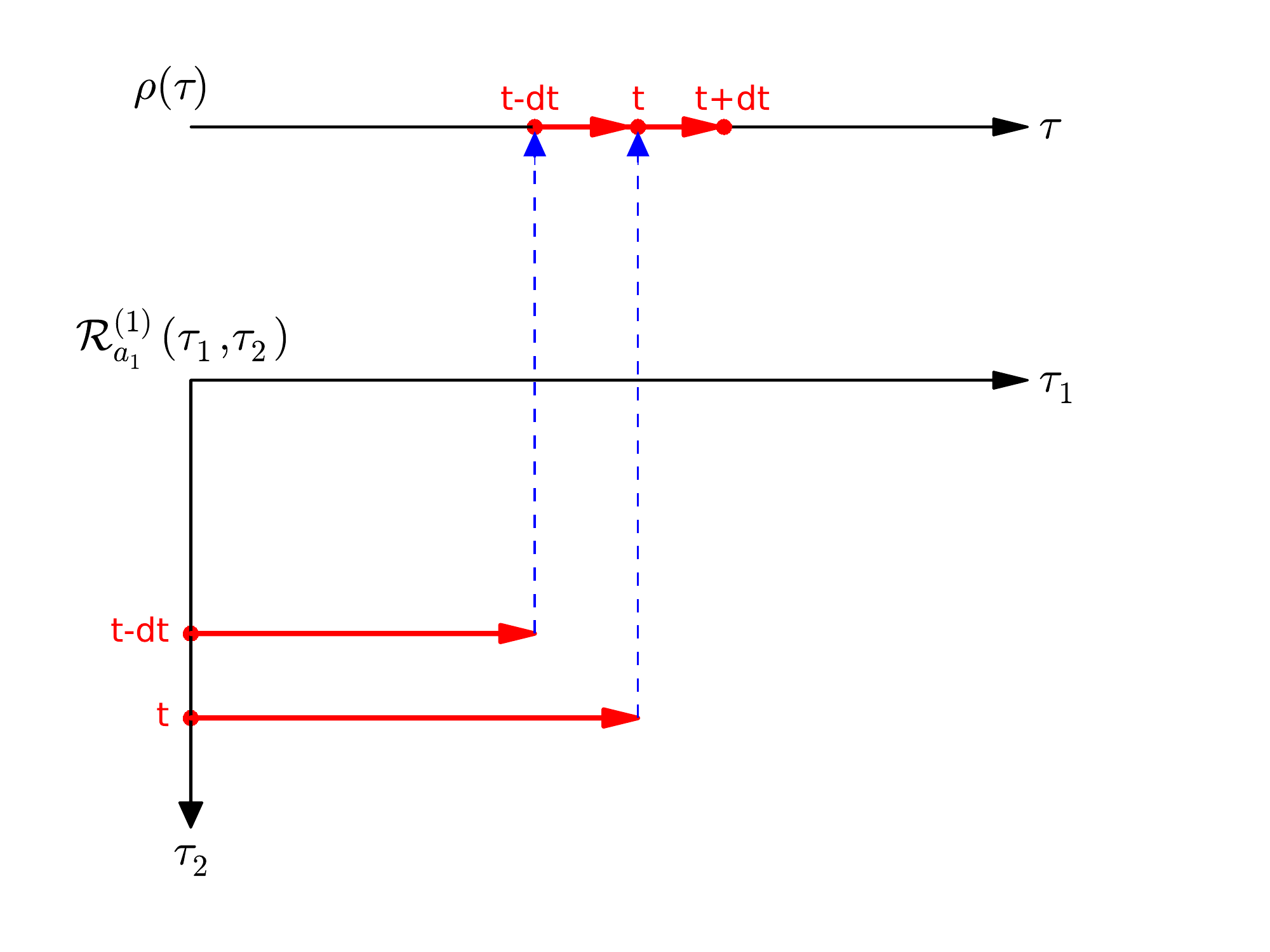}
		\caption{Propagation scheme for the RSHQME method including the $0$th and the $1$st-tier only. Red arrows visualize propagation of the first time argument, blue dashed arrows indicate the dependency of the propagation step of $\rho$ on the $1$st-tier auxiliary density operator.}\label{fig:propagation_scheme}
	\end{figure}
	In order to provide some more insight, we explicitly consider the simplistic case of a noninteracting system attached to leads described in the wide-band limit, which can be calculated using only the density matrix and the $1$st-tier auxiliary density matrices. The corresponding propagation scheme is visualized in Fig.\ \ref{fig:propagation_scheme} for two infinitesimal time steps centered around time $t$. 
	In order to propagate $\rho$ from $t-dt$ to $t$, which is highlighted by a red arrow, we need the $1$st-tier auxiliary operators $\mathcal{R}_{a_1}^{(1)}(t-dt, t-dt)$. These objects are calculated by setting the second time argument to $t-dt$ and propagating the first time argument from $0$ to $t-dt$, as indicated by a red arrow. Thereby, we implicitly assumed that the density matrix of previous times is known. In the next propagation step of the density matrix from $t$ to $t+dt$, the object $\mathcal{R}_{a_1}^{(1)}(t, t)$ needs to be known which is calculated by setting the second time argument to $t$ and then propagating the first time argument from $0$ to $t$.

\bibliography{Bib}

\end{document}